\documentclass[%
reprint,
superscriptaddress,
 amsmath,amssymb,
 aps,
 prx,
floatfix,
]{revtex4-2}

\usepackage{graphicx}
\usepackage{array}
\usepackage{amsmath,amsfonts,amssymb}
\usepackage[ansinew]{inputenc}
\usepackage{color}
\usepackage{calc}
\usepackage[normalem]{ulem}
\usepackage[separate-uncertainty = true,multi-part-units=single]{siunitx}
\usepackage{placeins} 

\newcommand*\didv{\mathrm{d} I/\mathrm{d} V}
\newcommand{\Figref}[1]{Fig.~\ref{#1}}

\begin{document}
\title{Fluorescent single-molecule STM probe}

\author{Niklas Friedrich} \email{n.friedrich@nanogune.eu}
    \affiliation{CIC nanoGUNE-BRTA, 20018 Donostia-San Sebasti\'an, Spain}

\author{Anna Ros\l awska}
    \affiliation{Universit\'e de Strasbourg, CNRS, IPCMS, UMR 7504, F-67000 Strasbourg, France}

\author{Xabier Arrieta}
    \affiliation{Center for Materials Physics (CSIC-UPV/EHU) and DIPC, Paseo Manuel de Lardizabal 5, Donostia - San Sebasti\'{a}n 20018, Spain}

\author{Michelangelo Romeo}
    \affiliation{Universit\'e de Strasbourg, CNRS, IPCMS, UMR 7504, F-67000 Strasbourg, France}

\author{Eric Le Moal}  
    \affiliation{Universit\'e Paris-Saclay, CNRS, Institut des Sciences Mol\'eculaires d'Orsay, 91405, Orsay, France}

\author{Fabrice Scheurer}
    \affiliation{Universit\'e de Strasbourg, CNRS, IPCMS, UMR 7504, F-67000 Strasbourg, France}

\author{Javier Aizpurua}
    \affiliation{Center for Materials Physics (CSIC-UPV/EHU) and DIPC, Paseo Manuel de Lardizabal 5, Donostia - San Sebasti\'{a}n 20018, Spain}

\author{Andrei G. Borisov} 
    \affiliation{Universit\'e Paris-Saclay, CNRS, Institut des Sciences Mol\'eculaires d'Orsay, 91405, Orsay, France}

\author{Tom{\'a}\v{s} Neuman} \email{neuman@fzu.cz}
    \affiliation{Universit\'e Paris-Saclay, CNRS, Institut des Sciences Mol\'eculaires d'Orsay, 91405, Orsay, France}
    \affiliation{Institute of Physics, Czech Academy of Sciences, Cukrovarnick\'{a} 10, 16200 Prague, Czech Republic}
    
\author{Guillaume Schull} \email{guillaume.schull@ipcms.unistra.fr}
    \affiliation{Universit\'e de Strasbourg, CNRS, IPCMS, UMR 7504, F-67000 Strasbourg, France}

\begin{abstract}
The plasmonic tip of a scanning tunnelling microscope (STM) is functionalized with a single fluorescent molecule and is scanned on a plasmonic substrate. The tunneling current flowing through the tip-molecule-substrate junction generates a narrow-line emission of light corresponding to the fluorescence of the negatively charged molecule suspended at the apex of the tip, \textit{i.e.}, the emission of the excited molecular anion (trion). The fluorescence of this molecular probe is recorded for tip-substrate nanocavities featuring  different plasmonic resonances, for different tip-substrate distances and applied bias voltages, and on different substrates. We demonstrate that the width of the emission peak can be used as a probe of the trion-plasmon coupling strength and that the energy of the emitted photons is governed by the molecule interactions with its electrostatic environment. Additionally, we theoretically elucidate why the direct contact of the suspended molecule with the metallic tip does not totally quench the radiative emission of the molecule. 
       
\end{abstract}

\date{\today}


\maketitle

\section*{Introduction}
Scanning probe microscopes (SPM) have revolutionized our perception of the atomic-scale world, providing topographic, electronic, magnetic, optical or mechanical information of a surface with ultimate spatial resolution. To address some specific properties of the probed sample, it proved advantageous to modify the chemical nature of the extremity of the scanning probe tip. Molecule-functionalized tips have played here a key role, enabling unprecedented resolution over the skeleton of molecules \citep{Temirov2008sthm, Gross2009} or addressing electrostatic \citep{Wagner2019, mallada_real-space_2021,tallarida_tip-enhanced_2017, lee_microscopy_2018}, magnetic \citep{verlhac_atomic-scale_2019} and transport properties \citep{Schull2009, Lafferentz2009, Schull2011}. In parallel, tip-induced electroluminescence \citep{Qiu2003Vibrationally,zhang2016visualizing,imada2016real, doppagne2020single, roslawska2022prx, Dolezal2022Real}, photoluminescence \citep{yang_sub-nanometre_2020,imada_single-molecule_2021,roslawska2023} or Raman spectroscopies \citep{Zhang2013,Lee2019} have reached sub-nanometer spatial resolution by making use of the extreme confinement of electromagnetic fields at the end of scanning tunneling microscope (STM) tips made of plasmonic materials. These techniques have made it possible to measure the fluorescence and scattering properties of individual molecules lying flat on thin decoupling layers with sub-molecular precision. Transferring a fluorescent molecule to the tip of an SPM would allow sensing and mapping the local electromagnetic field of a sample with the same spatial resolution. This has been attempted with terylene molecules embedded in micron-size nanocrystals \cite{Michaelis2000}, semiconductor quantum dots (SCQD) fixed on the tip of an optical fiber \cite{Cadeddu2016}, and NV centers in diamond nanocrystals attached to atomic force microscope tips \cite{Rondin2014}. Whereas these nearfield probes have been successfully used to sense electrostatic, electromagnetic and magnetic fields, the spatial resolution they provide remains limited by the (large) size of the emitter (SCQD) or of the nanocrystal into which the emitter is embedded. Hence, SPM tips preserving the fluorescence properties of a single-molecule directly attached to the tip apex are highly desirable, but have not been reported so far. This is presumably because direct contact between the molecule and the metal generally causes total quenching of the molecule fluorescence, due to ultrafast electron and energy transfer from the excited molecule to the metal \citep{doppagne2018electrofluorochromism}. 

Here we build on recent works demonstrating that a single 3,4,9,10-perylenetetracarboxylic-dianhydride (PTCDA) molecule can be suspended at the apex of a metallic STM tip \cite{Wagner2015sqdm, temirov2018molecularmodel}, and report on the electrically-induced fluorescence of this functionalized PTCDA tip. By performing a theoretical analysis, we show that in this configuration the spatial overlap between the molecular and tip electronic orbitals is much smaller than in geometries where the molecule would lie flat on the substrate. This leads to a reduced charge-transfer rate between the molecule and the metal, thus preserving the molecular fluorescence properties. As the PTCDA molecule is hanging approximately aligned with the tip axis in the STM junction, the molecular transition dipole is close to collinear with the plasmonic electric field in the gap, a situation that strongly contrasts with the on-surface flat-lying configuration of the molecules studied in usual tip-induced fluorescence measurements \cite{imada_single-molecule_2021}. This collinear geometry leads to a large increase of the coupling strength between the molecular exciton and the tip-sample plasmonic cavity by up to two orders of magnitude compared to the perpendicular (flat-lying) configuration. Eventually, the spectral characteristics of the PTCDA tips are investigated as a function of the plasmonic response of the tip-sample cavity, the bias voltage and the tip-sample distance, revealing that the emission of PTCDA tips can be used as a probe of the local electromagnetic and electrostatic fields. 

\noindent  

\begin{figure}
  \includegraphics[width=0.9\linewidth]{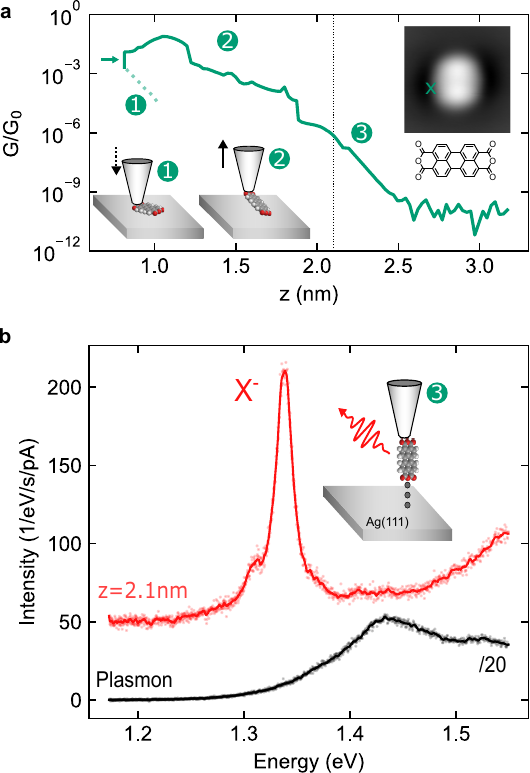}
  \caption{\label{fig1}
  \textbf{a} Conductance $G$ of the junction as a function of tip-substrate distance $z$ recorded during the PTCDA attachment procedure in units of the conductance quantum $G_0$. The tip is approached towards the molecule (dashed line, (1)) until a contact forms, indicated by the sudden increase of conductance (green arrow). Upon retracting the tip, the molecule is first cleaved from the substrate (2) and finally completely detached from the substrate (3). The minor deviation from perfect exponential decay of conductance at $\sim \SI{2.15}{nm}$ is caused by a change of voltage from $\SI{20}{mV}$ to $\SI{2.5}{V}$ during the retraction procedure. The inset is an STM constant current image ($I=\SI{30}{pA}$, $V=\SI{-300}{mV}$, $4\times \SI{4}{nm^2}$) of the molecule before lifting, as well as its Lewis structure. The position where the tip is approached is indicated by a green cross.
  \textbf{b} STML spectra of a metallic (black) and a PTCDA tip (red) in front of a bare Ag(111) surface. The peak at $\approx \SI{1.34}{eV}$ is associated to the trion fluorescence (X$^-$) of the suspended PTCDA.
  X$^-$ acquisition parameters: $V = \SI{2.5}{V}$, $I= \SI{70}{pA}$, $t=\SI{6}{min}$. Plasmon acquisition parameters: $V = \SI{2.5}{V}$, $I= \SI{50}{pA}$, $t=\SI{2}{min}$.
  } 
\end{figure} 

\section*{Results}
The procedure for making fluorescent PTCDA tips in a low-temperature, ultra high vacuum STM is schematized in Figure \ref{fig1} (see Methods for details of the sample preparation). A silver-covered tungsten tip is approached (dotted line in \Figref{fig1}\textbf{a}) to the oxygen atoms located at the extremity of a PTCDA molecule deposited on Ag(111) until a contact is reached (current jump indicated by the green arrow in \Figref{fig1}\textbf{a}). At this contact point, the last tip atom is expected to be at $z=\SI{0.8}{nm}$ from the metallic surface \citep{Fournier2011}. The tip is then retracted together with the attached molecule at its extremity. During this procedure the electrical current traversing the junction is monitored (solid green line \Figref{fig1}\textbf{a}). The presence of the molecule at the tip is confirmed by the much larger conductance values recorded during retraction ((2) \Figref{fig1}\textbf{a}).

Eventually, the complete detachment of the molecule from the surface is identified ((3) \Figref{fig1}\textbf{a}) at $z > \SI{2}{nm}$ by a strong slope change of the exponentially decaying conductance with distance. As discussed in Ref.~[\onlinecite{Knol2021}] and [\onlinecite{arefi_design_2022}], the PTCDA molecule adopts an unexpected up-standing configuration at the tip apex thanks to stabilizing electrostatic dipole forces that prevent the molecule from toppling over onto the tip shaft. Besides, it has been demonstrated that for low positive voltage tunneling conditions ($V < \SI{3}{V}$), the attached molecule is negatively charged (PTCDA$^-$) \cite{Temirov2008, Wagner2015sqdm, temirov2018molecularmodel, Esat2018, Esat2023}, a state that reflects the electron acceptor character of the molecule. In \Figref{fig1}\textbf{b} we report STM-induced luminescence (STML) spectra acquired at $V = \SI{2.5}{V}$, first with a clean metal tip, revealing a broad feature characteristic of plasmonic emission (black spectrum), and then with a fully detached ($z = \SI{2.1}{nm}$) PTCDA tip on top of the bare silver surface (red spectrum). In contrast to the spectrally broad plasmonic emission observed when the clean tip is used, the STML spectrum measured using the PTCDA tip exhibits a much sharper and more intense emission peak, which is centered at a photon energy of $\SI{1.34}{eV}$. In agreement with a recent report \cite{Dolezal2022}, we assign this peak to the luminescence of negatively charged PTCDA (later referred to as negative trion X$^-$). This result provides experimental evidence that the luminescence properties of the molecule suspended at the tip are preserved, despite the direct molecule-metal contact. The PTCDA tip can therefore be considered as a fluorescent molecular probe, whose spectral emission properties measured in the far field can be used to probe its local environment.

\begin{figure*}
  \includegraphics[width=1\linewidth]{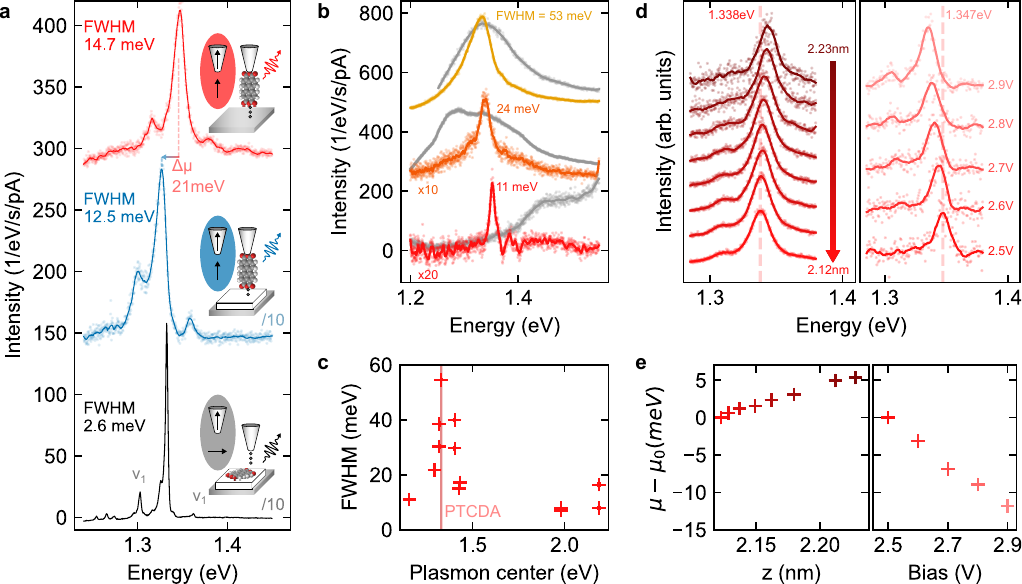}
  \caption{\label{fig2} 
  \textbf{a} STML spectra of a PTCDA tip in front of Ag(111) (red) and of 2ML NaCl (blue), and of the flat-lying PTCDA on 3ML NaCl excited with a metal tip (black). This data set was recorded with the same molecule. The plasmonic response of the tip (without molecule attached) is presented in SI \citep{SM} showing a very broad emission around $1.4 - \SI{1.5}{eV}$.
  \textbf{b} X$^-$ emission spectra (yellow, orange and red) for three different PTCDA tips above Ag(111) and the associated plasmonic spectra (in grey) recorded prior to the molecular functionalization of the tip. The FWHM of the X$^-$ peak increases for resonant plasmon-trion conditions.
  \textbf{c} FWHM of X$^-$ as a function of the energy of the plasmon peak maximum. These data were recorded at an identical set point ($I=\SI{30}{pA}$, $V=\SI{2.5}{V}$) and stems from 13 macro- or microscopically different tips.
  \textbf{d} X$^-$ emission spectra for varying tip-substrate separation (left panel, $V=\SI{2.5}{V}$) and for varying bias voltage (right panel, $z=\SI{2.2}{nm}$). A red shift of the central emission is observed upon reducing $z$ or increasing $V$.
  \textbf{e} Relative shift of the central emission line $\mu$ extracted from the data in \textbf{d}, taking the reference energy $\mu_0$ from the bottom spectrum in \textbf{d}.
  Spectra in panels \textbf{a}, \textbf{b} and \textbf{d} are offset vertically for easier comparison. Both $\mu$ and the $\text{FWHM}$ of the spectra were extracted by fitting a Lorentzian function to the data, accounting for the vibronic satellites with Lorentzian functions when necessary.
  Acquisition parameters of \textbf{a} top to bottom: $V = 2.5,\ 2.5,\ -\SI{2.5}{V}$, $I= 60,\ 50,\ \SI{60}{pA}$, $t=6,\ 6,\ \SI{30}{min}$. Spectra acquisition parameters for panel \textbf{b} - \textbf{d} are detailed in SI \cite{SM}.
  }
\end{figure*}

As a first step to characterize this fluorescent probe, we compare spectra (\Figref{fig2}\textbf{a}) successively obtained with a PTCDA tip located above the bare Ag(111) (red spectrum), above a two atomic layer-thick (2 ML) NaCl island grown on Ag(111) (blue spectrum) and with the exact same molecule adsorbed flat on 3ML NaCl but excited using a bare metal tip (black spectrum). For this last spectrum, the molecule was released from the tip by applying a low voltage pulse on top of NaCl. All spectra exhibit an intense X$^-$ emission line together with two vibronic satellites ($\nu_1$), located at $\approx \SI{30}{meV}$ on each side of the X$^-$ line, that occur from an in-plane breathing mode of the molecule \cite{Paulheim2016}. When the PTCDA tip is used, we observe that the STML spectrum measured on top of 2\,ML NaCl-covered Ag(111) is rigidly redshifted by $\Delta \mu = \SI{21}{meV}$ relative to the spectrum measured on bare Ag(111). This shift suggests that the emission properties of the tip-suspended molecule are dependent on its local dielectric environment, which is modified by the presence of the NaCl island. 

Moreover, the full width at half maximum (FWHM) of the emission peak is one order of magnitude larger in the spectra measured using the PTCDA tip  ($\text{FWHM} > \SI{12}{meV}$) than in the spectra measured using the clean tip on the PTCDA molecule lying flat on the NaCl ($\text{FWHM} \approx \SI{2.6}{meV}$). The width of molecular emission lines in STML experiments may have different origins and may be due to dephasing induced by the coupling of the trion with its electrostatic and/or phononic environments, non-radiative decay paths or a strongly shortened trion fluorescence lifetime compared to the free molecule because of the particular electromagnetic environment of the STM junction (Purcell effect). When the latter dominates, the FWHM is determined by the trion-plasmon coupling strength.

To identify the respective role of dephasing and lifetime-shortening on the emission linewidth of the PTCDA tip, we investigate the FWHM of the X$^-$ line as a function of the energy of the gap plasmon of the tip-sample junction. It is well known that the plasmonic response of the junction strongly depends on the nanoscale structure of the tip apex, a parameter that can be tuned by controlled indentations of the tip in the bare metal surface \cite{imada2017}. In \Figref{fig2}\textbf{b}, we show the STML spectra of three different PTCDA tips above bare Ag(111) together with the plasmonic response of the tip-sample junction recorded before functionalization with the PTCDA molecule. The plasmon peak is either strongly blue-detuned (bottom spectrum), weakly red-detuned (middle spectrum), or in-resonance (top spectrum) with the X$^-$ line. Here, the smaller the frequency mismatch between the plasmonic resonance and the trion emission, the broader the FWHM of the X$^-$ emission. This correlation is better evidenced in \Figref{fig2}\textbf{c} where the FWHM of the X$^-$ line is plotted as a function of the energy of the plasmon resonance maximum for 13 different tips. This plot also reveals that the FWHM does not reach values lower than $\SI{7}{meV}$, even for extremely non-resonant conditions. This suggests that the line width is in this case limited by dephasing or non-radiative decay paths that do not involve coupling to plasmons. In contrast, the FWHM increase observed at resonance indicates a reduction of the excited state lifetime due to the stronger coupling of the trion with the gap plasmons. The FWHM then reaches values as high as $\SI{53}{meV}$, corresponding to a fluorescence lifetime of $\SI{75}{fs}$. The trion-plasmon coupling is here still below the threshold of strong coupling \cite{chikkaraddy2016single}. As a figure of merit to quantify the regime of plasmon-trion coupling, we use the ratio between the experimental plasmon-trion coupling $g$ (\textit{i.e.}, the Jaynes-Cummings coupling strength as defined in Ref.~[\onlinecite{roslawska2022prx}]) and the experimental width of the plasmon $\kappa$. Assuming that the trion couples to a single plasmon mode as described in Ref.~[\onlinecite{roslawska2022prx}], we estimate $g$ for the resonant case (top spectrum \Figref{fig2}\textbf{b}) from the plasmon-induced width $\hbar\gamma_{\rm pl}\approx \SI{53}{meV}$ and $\hbar\kappa\sim \SI{100}{meV}$  ($\hbar$ is the reduced Planck constant) as $\hbar g=\hbar \sqrt{\gamma_{\rm pl}\kappa}/2\approx \SI{36}{meV}$. Since $2g$ is smaller than $\kappa$ we conclude that the plasmonic losses still dominate over the plasmon-trion coupling and we therefore observe a weak trion-plasmon coupling. For flat PTCDA adsorbed on 4 ML NaCl, a FWHM as low as $\SI{0.5}{meV}$ could be measured (see Supporting Information (SI) \citep{SM}), setting a lower limit for the trion lifetime of $\SI{8}{ps}$ when the tip and molecular dipoles are orthogonal to each other (compare sketches in Fig.\ref{fig2}\textbf{a}).

Overall, our results show that the excited state lifetime of the PTCDA trion is two orders of magnitude shorter when the transition dipole of the molecule is parallel to the electric field of the picocavity plasmons than when it is orthogonal to it, \textit{i.e.}, than in the configuration used in previous STML studies. Furthermore, the coupling strength is maximum when the plasmon is tuned in resonance with the X$^-$ line. This also indicates that the FWHM of X$^-$ can be used to probe variations of the local density of photonic states at metallic surfaces, \textit{e.g.}, at the surface of plasmonic nanostructures \cite{Michaelis2000, schell2014}.

In principle, not only the FWHM, but also the energy position of the X$^-$ line can be used for probing the environment of the molecular-tip. In previous reports, emission peak positions have been shown to vary as a function of both the local electromagnetic field, an effect known as photonic Lamb shift \cite{zhang2017sub}, and the electrostatic field where the spectral shift is described in terms of a Stark effect \citep{roslawska2022prx}. In \Figref{fig2}\textbf{d}, we monitor the X$^-$ line as a function of the tip-sample distance (at constant voltage) and as a function of the voltage (at constant tip-sample distance). The FWHM of the line ($\sim \SI{22}{meV}$) remains constant for the two sets, indicating that the trion-plasmon coupling does not significantly vary over this range of distances or voltages. In contrast, the X$^-$ line experiences a shift to lower energies with tip approach and with increasing voltages. \Figref{fig2}\textbf{e} reveals that this line-shift evolves essentially linearly with distance, at a rate of $\sim \SI{5}{meV/\AA}$, and with voltage, at a rate of $\sim \SI{30}{meV/V}$. These values are close to the rates reported for phthalocyanine molecules lying flat on NaCl and addressed by a metal tip \citep{roslawska2022prx}. In this case, the line-shifts caused by changes in the tip-sample distance and voltage were dominated by the Stark effect due to the electrostatic field in the STM junction. This effect is likely at play in \Figref{fig2}(d,\,e), suggesting that one can use the spectral shift of the PTCDA tip emission line as a probe of the local electrostatic field.

\section*{Discussion}
To explain the unexpected observation of a bright fluorescence from a molecule that is in direct contact with a metallic electrode, we perform a series of quantum chemical and quasi-electrostatic  calculations. We develop a theoretical description that compares (i) the interaction between the molecular exciton and the tip/substrate plasmons as discussed in detail in Ref.~[\onlinecite{roslawska2022prx}], and (ii) the charge transfer between the molecule and the tip, which is expected to strongly contribute to the quenching of the molecule's photon emission. Our model is thus able to capture the most relevant physical processes providing order-of-magnitude estimates of the photon emission efficiency.

\begin{figure}
  \includegraphics[width=1.0\linewidth]{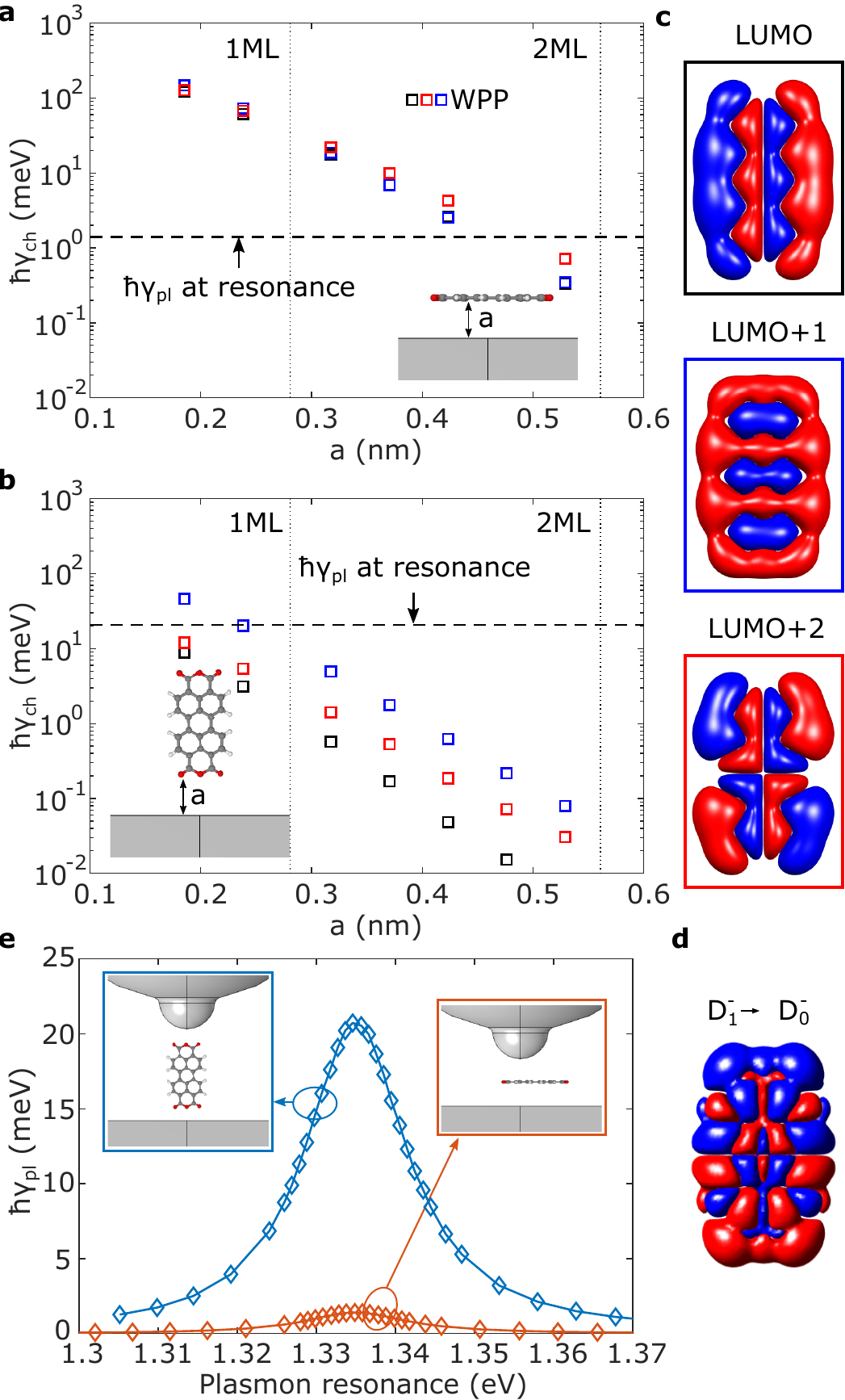}
  \caption{\label{fig3}
  \textbf{a}, \textbf{b} Calculated charge transfer rates $\gamma_{\rm ch}$ for \textbf{a} the molecule lying flat and \textbf{b} standing upright on a semi-infinite jellium interface as a function of the distance $a$ of the molecule from the jellium edge. The rates were calculated using the real-time wavepacket propagation method (WPP, squares) for orbitals shown in \textbf{c}. The color of the frames in \textbf{c} corresponds to the color coding in \textbf{a}, \textbf{b}. The dashed line in \textbf{a}, \textbf{b} corresponds to the maximum plasmon-induced decay rate $\gamma_{\rm pl}$ shown in \textbf{e} for the respective configurations. \textbf{c} Molecular orbitals LUMO, LUMO+1, and LUMO+2 calculated using Octopus. 
  \textbf{d} Transition charge density of the D$_1^-$ $\to$ D$_0^-$ transition calculated using TDDFT. 
  \textbf{e} Plasmon-induced exciton decay rate as a function of the plasmonic resonance energy tuned by varying the length of the model plasmonic cavity (see SI \citep{SM} for details of the geometry \citep{SM}) calculated for the molecule in the upright configuration (blue) and the flat configuration (orange). The molecule is positioned in the middle of the tip-substrate gap which is \SI{2.1}{nm} for the upright configuration and \SI{1.1}{nm} for the flat one.} 
\end{figure}

We describe the charge transfer in an effective single-particle picture in which we estimate the charge transfer rate of several molecular orbitals into the metal surface using the real-time wave-packet propagation (WPP) technique \citep{Aguilar2021} (see SI \citep{SM} for details and additional estimates using a perturbation approach).
The molecule is assumed to be positioned above a planar metal surface and the distance $a$ between the molecule and the image plane of the surface is varied. We consider two different configurations: the molecule lying flat on the surface (\Figref{fig3}\textbf{a}) and the molecule in the lifted configuration with its plane perpendicular to the surface (\Figref{fig3}\textbf{b}). In the latter case, $a$ describes the distance between image plane and the bottom oxygen atoms of the molecule.
The decay rate $\gamma_{\rm ch}$ induced by the charge transfer is shown in \Figref{fig3}\textbf{a} (\Figref{fig3}\textbf{b}) for the flat-lying (lifted) configuration for the LUMO (lowest unoccupied molecular orbital) (black), LUMO+1 (blue), and LUMO+2 (red) as squares calculated using the full real-time WPP. The orbital labeling is derived from the electronic configuration of the neutral molecule, \textit{i.e.}, the LUMO is split into the singly occupied and unoccupied orbitals in the negative ground state D$_0^-$. We note that both LUMO and LUMO+2 have a significant contribution to the electronic configuration of ${\rm D}_1^-$ (see SI \citep{SM} for details). 
The corresponding orbitals are shown in \Figref{fig3}\textbf{c}. \Figref{fig3}\textbf{b} (\Figref{fig3}\textbf{a}) demonstrates that at a distance of about $\SI{0.2}{nm}$ - roughly corresponding to the distance between the atoms binding the tip and the molecule \cite{Knol2021, arefi_design_2022} - the electronic decay rate of the orbitals reaches up to $\sim \SI{10}{meV}$ ($\sim \SI{100}{meV}$) for the lifted (flat-lying) configuration. In other words, the lifted configuration leads to a ten-times weaker metal-molecule electronic coupling than for a flat-lying molecule.

To know if the charge transfer rate $\gamma_{\rm ch}$ is fast enough to quench the trion emission, one should compare it with the radiative decay rate expressed by $\gamma_{\rm pl}$. To calculate $\gamma_{\rm pl}$, we first describe the electronic excitations in the negative PTCDA molecule using time-dependent density-functional theory (TDDFT) (see SI \citep{SM} for details). We relax the molecular geometry in its first excited state D$_1^-$ and extract the molecule's transition charge density $\rho$ (\Figref{fig3}\textbf{d}) - generalizing the concept of the transition dipole moment. We next calculate a quasi-static electric potential $\phi$ generated by the source charge density $\rho$ placed in plasmonic environment formed by the tip and the substrate. We obtain the plasmon-induced broadening $\hbar\gamma_{\rm pl}$ of the trion as $\hbar\gamma_{\rm pl}=2{\rm Im}\{\int \rho({\bf r})\phi({\bf r}){\rm d}{\bf r} \}$ \cite{roslawska2022prx} and show the result in \Figref{fig3}\textbf{e} as a function of the plasmon resonance energy, which is artificially tuned by changing the tip geometry (see SI \citep{SM}). The molecule is considered in two geometrical configurations in the center of the tip-substrate gap, once parallel to the surface, and once in the upright geometry. As demonstrated in \Figref{fig3}\textbf{e}, the resonance between the trionic state and the plasmonic mode is necessary to reach large trion broadening of up to $\SI{21}{meV}$ for the upright molecule, on the order of the experimental value reported in \Figref{fig2}\textbf{c} (see SI \citep{SM} for a detailed discussion). Our calculation thus supports the conclusion that the plasmon resonance is responsible for the observed broadening of the trion line. Besides, in the lifted configuration, the broadening is $\approx 15$ times larger than the maximal plasmon-exciton coupling obtained for the flat-lying molecule (\Figref{fig3}\textbf{e}). In other words, the fluorescence decay is $\approx 15$ times faster for molecules oriented aligned with the tip axis than for flat-lying molecules. The plasmon-induced trion decay rate $\gamma_{\rm pl}$ in both configuration is represented by horizontal dashed line in \Figref{fig3}\textbf{a}, \textbf{b}. For the flat-lying case (\Figref{fig3}\textbf{a}), we see that the plasmon-induced trionic decay rate dominates over the charge-transfer rate $\gamma_{\rm ch}$ only for $a \gtrapprox \SI{0.45}{nm}$, a molecule-substrate distance that is close to the one corresponding to 2 ML NaCl where STM-induced fluorescence signal starts to be observed. In contrast, the plasmon-induced trion decay rate already dominates for $a \gtrapprox \SI{0.2}{nm}$ for the lifted configuration (\Figref{fig3}\textbf{b}). Overall, these data show that the observation of fluorescence in the lifted configuration results from the combination of a reduced electronic coupling between molecule and electrode states and from the enhanced electromagnetic coupling between the vertically polarized tip-plasmons and the molecular excited states.  

\begin{figure}
  \includegraphics[width=1\linewidth]{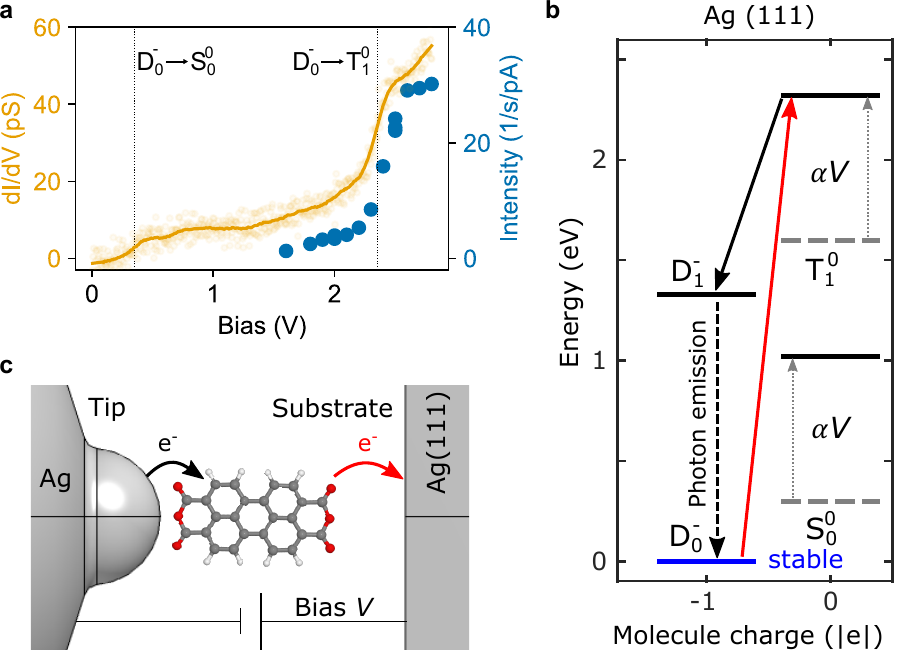}
  \caption{\label{fig4}
  \textbf{a} $\didv$ (yellow) and photon intensity (blue dots) shown as a function of sample bias. 
  \textbf{b} Tentative diagram of many-body electronic states of the molecule and the tip-mediated (black arrow) and substrate-mediated (red arrow) charge-transfer events followed by trion recombination (black dashed arrow). Due to the non-negligible voltage drop across the molecule, the relative energies of the states (aligned with respect to the Ag(111) work function) are dependent on the applied bias (see details in SI \citep{SM}) with $\alpha\approx 1/3$ (grey dotted arrows). The direct contact of the molecule with the tip stabilizes the molecule in the negative doublet state D$_0^-$, but when bias is applied, the molecule can be brought into the excited state D$_1^-$ by first releasing an electron into the substrate (red arrow) and then capturing an electron from the tip (black arrow) as discussed in the text.
  \textbf{c} Schematic depiction of the tip-molecule-substrate geometry where voltage is applied between the two electrodes and an electron is transferred from the tip to the molecule and subsequently from the molecule into the substrate Ag(111) surface.
  } 
\end{figure}

Next, we briefly discuss the mechanisms that are likely involved in the STM-induced excitation of the PTCDA molecule suspended on the tip. In STML the mechanism bringing the molecule to the excited state usually depends on a sequence of charge transfer events that are specific to the considered system \cite{miwa2019, jiang2023}. Figure \ref{fig4}\textbf{a} shows a $\didv$ spectrum simultaneously recorded with the total intensity of the X$^{-}$ line emitted by the molecule. The $\didv$ curve shows a smooth step-like increase of the bias voltage at $\sim \SI{0.3}{V}$ and $\sim \SI{2.3}{V}$. The step at $\sim \SI{2.3}{V}$ is accompanied by an equally smooth onset of photon emission. To explain this voltage-dependent behaviour we propose a model of charge transfer events between the many-body electronic states of the molecule (see \Figref{fig4}\textbf{b}). In the model we include electronic states of the neutral molecule (singlet ground state S$_0$ and a triplet state T$_1^0$), two doublet states (D$_0^-$ and D$_1^-$) of the negatively charged molecule. The energies of the states displayed in the diagram are aligned with respect to the work function of the Ag tip which we estimate at $W_{\rm Ag}\approx \SI{4.5}{eV}$ \cite{dweydari1975, CHULKOV1999330}. When a voltage $V$ is applied, the relative energy of the different charge states shifts by $\alpha V$ as indicated by the grey dotted arrows. Here $\alpha$ indicates the fractional voltage drop experienced by the molecule with respect to the tip electrode. We envision the mechanism as follows. First, by applying a small bias of about $\sim \SI{0.3}{V}$ the singly negatively charged molecule can be transiently neutralized by tunneling to the substrate and brought into the S$_0^0$ state. For a voltage of $V \approx \SI{2.3}{V}$ the transition into the neutral triplet T$_1^0$ state is enabled (red arrows in \Figref{fig4}\textbf{a}, \textbf{c}). Once in the neutral triplet state, the molecule can be rapidly charged from the tip (black arrows) and can end up in the excited doublet configuration (D$_1^-$). D$_1^-$ readily decays into the ground state either by non-radiatively exchanging an electron with the tip or by photon emission. We note that when the lifted molecule is brought above the NaCl/Ag(111) surface, the mechanism leading to light emission can differ from the present discussion (see SI \citep{SM}).

In conclusion, we demonstrated that a PTCDA molecule preserves its intrinsic emission properties even when it is \textit{directly} attached to a plasmonic scanning probe tip. This is due to the relatively low spatial overlap between the tip and molecule electronic orbitals --- which in turn leads to weak luminescence quenching by charge transfer --- and by the strongly increased radiative decay probability (by up to 2 orders of magnitude compared to flat-lying PTCDA) of the molecular emitter fixed vertically at the apex of the plasmonic tip. This increased trion-plasmon interaction, however, remains insufficient to reach the strong coupling regime. Our data also demonstrate that the fluorescent properties of the molecular probe are sensitive to their electromagnetic and electrostatic environment. Eventually, the excitation mechanism by tunneling electrons has been discussed. Further work will be devoted to characterizing the spatial resolution that is achievable with this atomic-scale sensor that could be used, in the close future, in resonant energy transfer microscopy experiments having ultimate lateral and axial precision.         

\section*{Methods}
The STM data was acquired with a low temperature ($\SI{5.5}{K}$) Unisoku setup operating in ultrahigh vacuum and adapted to detect the light emitted at the tip-sample junction. The optical detection setup is composed of a spectrograph coupled to a CCD camera; a grating having a groove density of 300 lines/mm was used and provided a spectral resolution of $\approx \SI{1}{nm}$ for all the data presented in the paper with the exception of the data of Fig. S3 where a grating with 1200 lines/mm was used. Tungsten STM-tips were introduced in the sample to cover them with silver so as to tune their plasmonic response. The Ag(111) substrate was cleaned with successive Ar$^+$-ion sputtering and annealing cycles.
 
Approximately 0.5 monolayers of NaCl were sublimated on Ag(111) kept at room temperature. The sample was then flash-annealed up to $\SI{370}{K}$ to obtain square domains of bi- and tri-layers of NaCl. PTCDA was evaporated \textit{in situ} on the sample held at $\SI{5.5}{K}$ using a molecular beam evaporator ($T_\text{evap}=\SI{613}{K}$), resulting in a sparse distribution of individual molecules.

Differential conductance spectra were recorded with internal lock-in amplifier using a modulation amplitude of $\SI{20}{mV}$ at $f=\SI{768}{Hz}$. Light and $\didv$ spectra were processed using custom Python scripts. The Python package \texttt{lmfit} \citep{lmfit} was used for all fitting procedures. STM images were processed using WSXM \citep{horcas_wsxm_2007}.

\section*{References}

\section*{Acknowledgements}
JA and XA would like acknowledge the Spanish Ministry of Science and Innovation, Project Nr. PID2022-139579NB-I00, and XA acknowledges his grant Nr. FPU21/02963. 
NF acknowledges funding from the Spanish government MCIN/AEI/10.13039/501100011033 through grants MAT2016-78293-C61 and  PID2019-107338RB-C61, and by the European Union (EU) H2020 program through the FET Open project SPRING (grant agreement No. 863098). This work is supported by ``Investissements d'Avenir'' LabEx PALM (ANR-10-LABX-0039-PALM). TN acknowledges the Lumina Quaeruntur fellowship of the Czech Academy of Sciences. Computational resources were supplied by the project "e-Infrastruktura CZ" (e-INFRA CZ LM2018140) supported by the Ministry of Education, Youth and Sports of the Czech Republic.
This project has received funding from the European
Research Council (ERC) under the European Union's Horizon 2020 research and innovation programme (grant agreement no. 771850). This work of the interdisciplinary Thematic Institute QMat, as part
of the ITI 2021 2028 program of the University of Strasbourg, CNRS and Inserm, was supported by IdEx Unistra (ANR 10 IDEX 0002), as well as by SFRI STRAT'US project (ANR 20 SFRI 0012) and EUR QMAT
ANR-17-EURE-0024 under the framework of the French investments for the Future Program. This work is supported by the Agence Nationale de la Recherche (ANR) under Contract No. ANR-22-CE09-0008.

\section*{Author information}
\subsection*{Contributions}
GS conceived and designed the experiment.
MR developed specialized software for the data acquisition.
NF, AR, EL performed the STML experiments.
NF analyzed the data.
XA, JA, AB and TN developed the theoretical models and performed the numerical simulations.
NF, TN and GS wrote the manuscript with input from all authors.

\section*{Ethics Information}
\subsection*{Competing interests}
The authors declare no competing interests.

\end{document}


\title{Supporting Information: Fluorescent single-molecule STM probe}

\author{Niklas Friedrich} \email{n.friedrich@nanogune.eu}
    \affiliation{CIC nanoGUNE-BRTA, 20018 Donostia-San Sebasti\'an, Spain}

\author{Anna Ros\l awska}
    \affiliation{Universit\'e de Strasbourg, CNRS, IPCMS, UMR 7504, F-67000 Strasbourg, France}

\author{Xabier Arrieta}
    \affiliation{Center for Materials Physics (CSIC-UPV/EHU) and DIPC, Paseo Manuel de Lardizabal 5, Donostia - San Sebasti\'{a}n 20018, Spain}

\author{Michelangelo Romeo}
    \affiliation{Universit\'e de Strasbourg, CNRS, IPCMS, UMR 7504, F-67000 Strasbourg, France}

\author{Eric Le Moal}  
    \affiliation{Universit\'e Paris-Saclay, CNRS, Institut des Sciences Mol\'eculaires d'Orsay, 91405, Orsay, France}

\author{Fabrice Scheurer}
    \affiliation{Universit\'e de Strasbourg, CNRS, IPCMS, UMR 7504, F-67000 Strasbourg, France}

\author{Javier Aizpurua}
    \affiliation{Center for Materials Physics (CSIC-UPV/EHU) and DIPC, Paseo Manuel de Lardizabal 5, Donostia - San Sebasti\'{a}n 20018, Spain}

\author{Andrei G. Borisov} 
    \affiliation{Universit\'e Paris-Saclay, CNRS, Institut des Sciences Mol\'eculaires d'Orsay, 91405, Orsay, France}

\author{Tom{\'a}\v{s} Neuman} \email{neuman@fzu.cz}
    \affiliation{Universit\'e Paris-Saclay, CNRS, Institut des Sciences Mol\'eculaires d'Orsay, 91405, Orsay, France}
    \affiliation{Institute of Physics, Czech Academy of Sciences, Cukrovarnick\'{a} 10, 16200 Prague, Czech Republic}
    
\author{Guillaume Schull} \email{guillaume.schull@ipcms.unistra.fr}
    \affiliation{Universit\'e de Strasbourg, CNRS, IPCMS, UMR 7504, F-67000 Strasbourg, France}

\date{\today}


{
\let\clearpage\relax
\maketitle
}
\onecolumngrid
\tableofcontents 

\FloatBarrier
\newpage
\section{Experimental}

\subsection{Plasmon corresponding to FIG.\,2\textbf{a}}
\FloatBarrier
\vspace{-0.3cm}
\begin{figure}[h!]
  \includegraphics[width=0.7\linewidth]{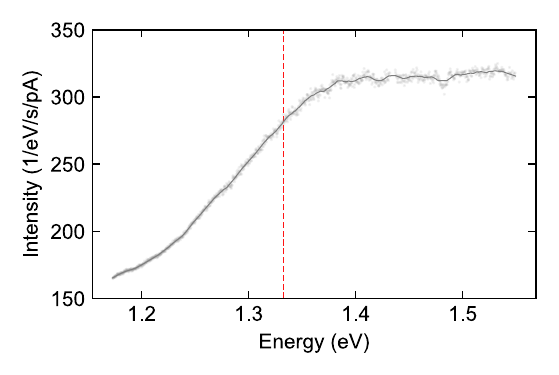}
  \vspace{-0.5cm}
  \caption{\label{sfig_plasmon}
  STML spectrum revealing the spectral response of the Ag tip- Ag substrate nanocavity plasmon used to collect the data presented in FIG.\,2\textbf{a}. The spectrum was acquired with $V=-\SI{2.5}{V}$, $I=\SI{100}{pA}$, $t=\SI{3}{min}$. The red dotted line indicates the emission energy of X$^-$.
  }
\end{figure}
\newpage
\FloatBarrier
\subsection{STML spectra on 4ML NaCl} 
\FloatBarrier
\vspace{-0.3cm}
\begin{figure}[h!]
  \includegraphics[width=0.7\linewidth]{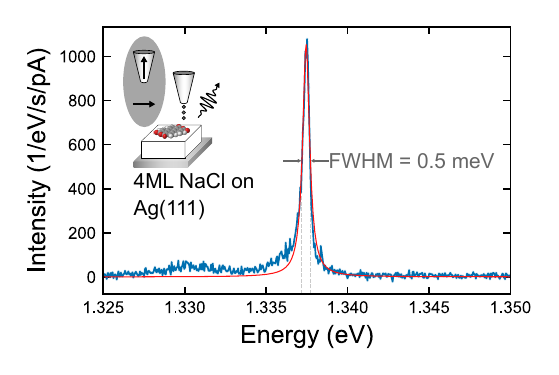}
  \vspace{-0.5cm}
  \caption{\label{sfig_4ML}
 STML spectrum of a PTCDA molecule adsorbed flat on 4ML NaCl/Ag(111) (blue curve) obtained with a 1200 lines/mm grating and a reduced spectrometer slit opening to optimize spectral resolution ($\approx \SI{0.22}{meV}$). The $\text{FWHM}=\SI{0.5}{meV}$ of the main resonance is extracted from a Lorentzian fit (red curve). The spectrum was acquired with $V=\SI{-2.5}{V}$, $I=\SI{50}{pA}$, $t=\SI{5}{min}$.
  }
\end{figure}
\newpage

\FloatBarrier
\subsection{Spectra acquisition parameters}

\begin{table}[h!]
    \centering
    \vspace{-0.5cm}
    \caption{Relevant acquisition parameters for spectra presented in FIG.\,2 of the main manuscript. All spectra were obtained above clean Ag(111). $X^-$ spectra were acquired with a PTCDA molecule attached to the tip, plasmon spectra with a clean metallic tip.}
    \footnotesize
    \begin{tabular}{l|c c c|l}
     FIG. \hspace{0.5cm} & \hspace{0.5cm} $V(\si{V})$ \hspace{0.5cm}  & \hspace{0.5cm} $I (\si{pA})$ \hspace{0.5cm} & \hspace{0.5cm} $t(\si{min})$ \hspace{0.5cm} & \hspace{0.5cm} Spectrum position in the figure \\ \hline \hline
     2\textbf{b}&  2.5 & 50 & 10 & top ($X^-$)\\
         &  2.5 & 50 &  6 & middle\\
         &  2.5 & 50 & 30 & bottom\\
         &      &    &    & \vspace{-0.2cm} \\
         &  2.5 & 50 &  2 & top (plasmon)\\
         &  2.5 & 1000 & 30 & middle\\
         &  2.5 & 100 & 6 & bottom\\
         &      &    &    & \vspace{-0.2cm} \\ \hline
     2\textbf{c}&  2.5 & 50 &  6 & all $X^-$ spectra \\
         &  2.5 & 50-1000   & 2-30   & all plasmon spectra, different $I$ and $t$\\ 
         &      &    &    & do not influence the plasmon central energy\\ 
         &      &    &    & \vspace{-0.2cm} \\\hline
     2\textbf{d}&  2.5 &  7 &  6 & top ($z$-approach)\\
         &  2.5 & 10 &  6 &\\
         &  2.5 & 20 &  6 &\\
         &  2.5 & 30 &  6 &\\
         &  2.5 & 40 &  6 &\\
         &  2.5 & 50 &  6 &\\
         &  2.5 & 60 &  6 &\\
         &  2.5 & 70 &  6 & bottom \\ 
         &      &    &    & \vspace{-0.2cm} \\\hline
     2\textbf{d}&  2.9 & 76 &  2 & top ($V$-dependency)\\
         &  2.8 & 66 &  2 &\\
         &  2.7 & 56 &  2 &\\
         &  2.6 & 40 &  2 &\\
         &  2.5 & 31 &  2 &  bottom \\ 
         &      &    &    & \vspace{-0.2cm} \\\hline
    \end{tabular}
    \normalsize
    \label{tab:my_label}
\end{table}
\FloatBarrier

\newpage
\section{Theory}
To interpret the experimental results, we develop a theoretical model addressing several important aspects of the tunneling process and light emission from PTCDA in an STM junction:

(i) We calculate the tunneling rates between the molecule and the respective electrodes (tip and substrate). To this end, we consider a single electron approximation, where an "active" molecular orbital is isolated within the many-body molecular state, and an electron tunneling between the molecule and the metal is seen as a one-electron energy-conserving transition between this orbital and electronic states of the metal (also termed as resonant electron transfer, RET). The electron dynamics in the system is described with the wave packet propagation (WPP) method \cite{WPP_Method2021} yielding the tunneling rates, energies of the "active" molecular orbitals in the STM junction, as well as the spatio-temporal picture of the process.

(ii) We calculate the plasmon-induced decay rate of the molecular excited state (trion) combining the classical electromagnetic description of the tip and the sample with a quantum description of the molecular exciton as described in Ref.[\onlinecite{roslawska2022prx}]. In brief, this model takes the molecular transition charge density (the generalization of molecular transition-dipole moment) obtained from linear-response TDDFT and inserts it as a classical source generating a quasi-static electric field in the dielectric environment formed by the tip and the substrate. 

\subsection{(Time-dependent) density-functional-theory calculations}
To model the exciton, we perform a linear-response time-dependent density-functional theory (TDDFT) calculation of the PTCDA molecule in a vacuum using Gaussian 16 \cite{g16} with the B3LYP hybrid functional \cite{becke1993b3lyp} and the augmented correlation-consistent valence double zeta basis set AUG-cc-pVDZ. This basis set is augmented by diffuse functions that we include to properly describe the spill-out of the electron density in negative ions. We relax the molecular geometry in its negative ground state D$_0^-$ assuming that the molecule is singly negatively charged when attached to the tip \cite{temirov2018molecularmodel, zonda2021}. The ground state of the negative molecule is modelled as a doublet state within the spin-unrestricted Kohn-Sham density functional theory (DFT). From TDDFT we obtain a series of excited states, the lowest of which, denoted D$_1^-$, appears at the excitation energy of $\SI{1.453}{eV}$. We next optimize the molecular geometry in D$_1^-$ and calculate the emission energy of the D$_1^-$$\to$D$_0^-$ transition $E_{{\rm D}_1^-\to {\rm D}_0^-}=\SI{1.335}{eV}$, in excellent agreement with the experiment. We also extract the transition charge density $\rho$ of D$_1^-$$\to$D$_0^-$ shown in FIG.\,3\textbf{d} of the main text, which we use as an input for further calculations of the plasmon-induced decay rate.

In order to describe charge transfer processes, we obtain the Kohn-Sham orbitals of the molecule from a DFT ground state calculation of a neutral PTCDA molecule using Octopus \cite{tancogne2020} in the local density approximation \cite{perdew1981} and using built-in atomic pseudopotentials. We use Octopus because it implements the Kohn-Sham scheme on a real-space grid, thus providing wave functions with correct asymptotic behaviour even at larger distances from the molecule. This is in contrast with the Gaussian representation of the orbitals in Gaussian 16 (and similar codes), which benefits from the efficiency of the basis sets, but cannot represent the molecular charge density (Kohn-Sham orbitals) far from the atoms.

\subsection{Exciton decay rate}

\begin{figure}
  \includegraphics[width=\linewidth]{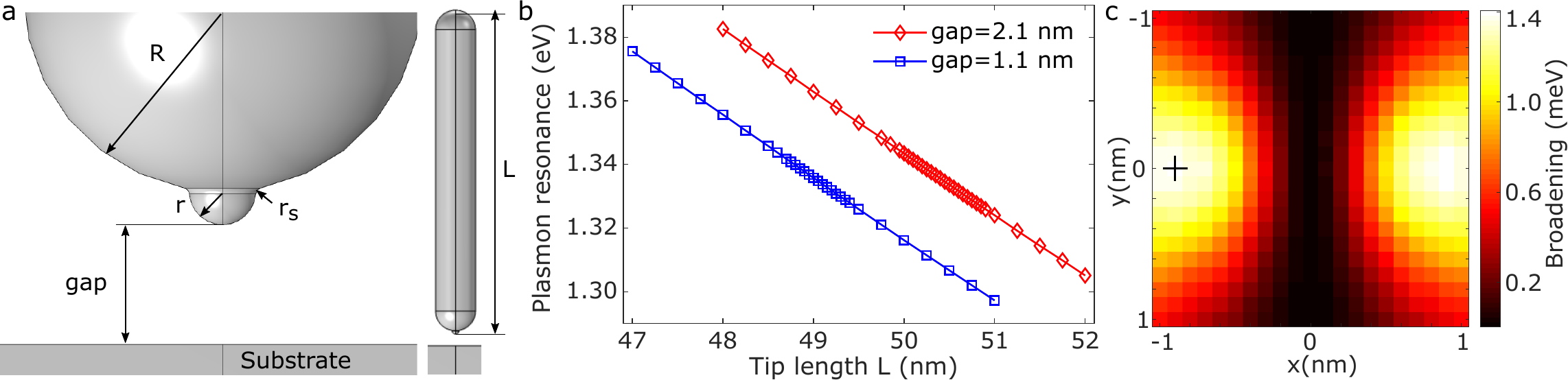}
  \caption{\label{fig:states} 
  \textbf{a} Geometry of the tip and the substrate used to model the plasmonic response. The following parameters were used for the calculation: $R=\SI{3}{nm}$, $r_{\rm S}=\SI{0.1}{nm}$, $r=\SI{0.5}{nm}$, gap=$\SI{2.1}{nm}$ (gap=$\SI{1.1}{nm}$) for the molecule aligned with (perpendicular to) the tip axis. $L$ was varied in the range shown in \textbf{b} to tune the plasmon resonance. 
  \textbf{b} Plasmon resonance as a function of the tip length for two different gap sizes (gap=$\SI{2.1}{nm}$ - red diamonds, and gap=$\SI{1.1}{nm}$ - blue squares) considered in this work. The position of the plasmon resonance was taken as the position of the maximum of the plasmon-induced broadening for the vertically oriented molecule calculated as a function of the exciton frequency that was varied as a free parameter. 
  \textbf{c} Calculated tip-position-dependent map of the line broadening for a molecule positioned perpendicularly to the tip axis in the middle of the gap. The position marked by the black cross is used to calculate the data in FIG.\,3\textbf{e} of the main text.}
\end{figure}
 
To estimate the plasmon-induced decay rate of the exciton shown in FIG.\,3\textbf{e} of the main text, we insert $\rho$ as a source charge density into a finite-element calculation of electrostatic field in the environment defined by the Ag electrodes: the STM tip and the substrate. The geometry of the tip and the substrate is shown in FIG.\,\ref{fig:states} and the relevant parameters are defined in the figure caption. The tip is modelled as a vertically oriented rod with hemispherical ends, one of which is facing the substrate. At the bottom of this hemispherical cap, there is an additional hemispherical protrusion which mimics the atomic sharpness of the STM tip. The substrate is modelled as a semi-infinite interface. The simulation is performed using the finite-element method implemented in Comsol Multiphysics, version 5.5 \cite{comsol}. We use the zero-potential condition on the domain boundary and enlarge the simulation domain until convergence is reached. 

From the simulation, we extract the quasi-static potential $\phi$ induced by the environment in response to the source density and estimate the decay rate $\gamma$ as $\hbar\gamma=2{\rm Im}\{\int \rho\phi {\rm d}{\bf r}\}$. We next tune the plasmon resonance by adjusting the length of the tip while maintaining the gap of $\SI{2.1}{nm}$ ($\SI{1.1}{nm}$) for the configuration of the molecule aligned with (perpendicular to) the tip apex. We probe the resonance by calculating the plasmon-induced broadening as a function of the exciton energy that is treated as a free parameter. The results of this calculation are shown in FIG.\,3\textbf{e} (main text) for the molecule aligned with the tip axis and perpendicular to it, respectively for several lengths of the tip. 

For the gap size of $\SI{2.1}{nm}$, and the center of the vertically oriented molecule positioned in the middle of the gap and the plasmon resonance tuned to the exciton energy, we obtain $\hbar\gamma\approx \SI{21}{meV}$ (see FIG.\,3\textbf{e} of the main text). We note that this estimate cannot exactly reproduce the value of the plasmon-enhanced decay of the exciton in the real junction as the exact geometry of the junction is generally not known and other factors, such as radiative damping, can also modify the plasmon response. We therefore treat the value of $\hbar\gamma\approx \SI{21}{meV}$ rather as an order-of-magnitude estimate of the plasmon-enhanced exciton decay rate that we compare to the charge-transfer rate calculated as described in section\,\ref{s:ctd}. 

We also calculate the plasmon-induced decay rate of the molecule in the configuration perpendicular to the tip axis shown in FIG.\,3\textbf{e} of the main text as a function of the tip-plasmon resonance. We position the molecule in the middle of the $\SI{1.1}{nm}$ gap and find the lateral position ($x,y$) of the tip that maximizes the interaction of the tip plasmon with the molecular exciton. To that end, we first calculate the tip-position-dependent map of the plasmon-induced broadening, shown in FIG.\,\ref{fig:states}\textbf{c}, maintaining the $\SI{1.1}{nm}$ gap and find the tip position for which the broadening is maximal (marked by the black cross in FIG.\,\ref{fig:states}\textbf{c}). 

\subsection{Charge transfer dynamics}\label{s:ctd}

\subsubsection{The wave packet propagation approach}\label{ss:WPP}

The WPP approach used here to describe the charge transfer dynamics between 
the molecule and the substrate was detailed earlier \cite{WPP_Method2021}. 
Therefore, only a brief discussion will be presented here. Within the 
one-electron picture the electron population of the molecular orbital active 
in RET decays via an energy-conserving electron transfer to the metal. Because of the coupling with the continuum (metal states), the discrete state (molecular orbital) becomes quasi-stationary. It is characterized by its energy and width (inverse of the lifetime), and appears as a resonance in electronic density of states of the system. Within the WPP approach the characteristics of the quasistationary molecule-localized electronic states are obtained from the analysis of the time-evolution of the electron wave function $\psi(\vec{r},t)$ described by the time-dependent Schr\"odinger equation (atomic units are used unless otherwise stated)

\begin{align}
    \label{eq:MainWPP}
    i\frac{\partial \psi(\vec{r},t)}{\partial t} 
    =\left[T+V_{\rm eff}(\vec{r})\right] \psi(\vec{r},t),
\end{align}

\begin{figure}
  \includegraphics[width=1.0\linewidth]{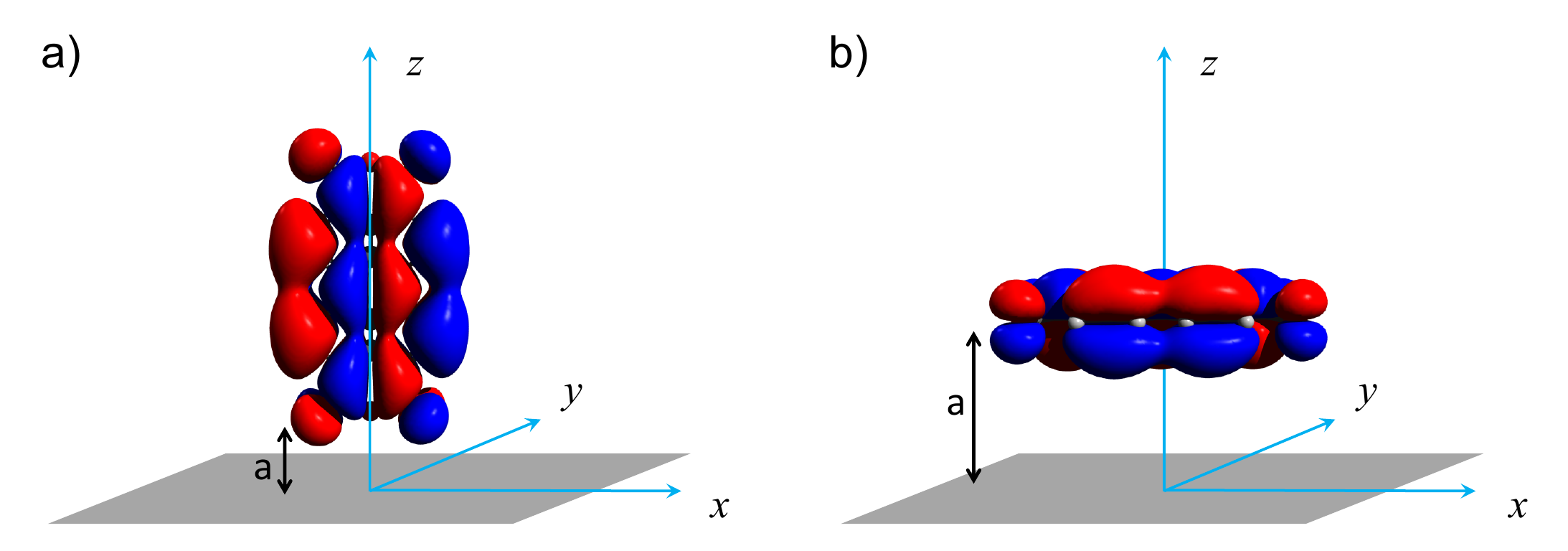}
  \caption{\label{fig:WPPgeometry} 
  Geometry of the molecule above the metal substrate used in 
  WPP calculations. The molecule represented by the wave function of the LUMO is above the metal surface in both the configuration \textbf{a} perpendicular to the surface and \textbf{b} parallel to the surface. 
  The latter configuration mimics the lifted molecule standing upright on the Ag tip.
  The molecule-surface distance 
  $a$ is measured from the image plane of the metal set as $(x,y,z=0)$-plane and shaded with gray color.
  }
\end{figure}

The initial conditions $\psi({\bf r},t=0)$ are given by the 
orbitals of the free-standing molecule identified as orbitals active 
in RET. In Eq.~\ref{eq:MainWPP}, the kinetic energy operator $T=-\frac{1}{2}\nabla^2$, 
and $V_{\rm eff}({\bf r})$ is an effective one-electron potential. 
It is defined assuming that the electron-molecule interaction is fully 
screened inside the metal.
\\
\begin{align}
\label{eq:EffPot}
    V_{\rm eff}({\bf r})=\begin{cases} 
    V_{\rm{mol}}(\vec{r})+V_{\rm surf}(z)+
\Delta V_{\rm surf}(\vec{r})+V_{\rm{abs}}(\vec{r}), &z\geq z_0, \\ 
    V_{\rm surf}(z)+V_{\rm{abs}}(\vec{r}), &z< z_0. \end{cases}
\end{align}
\\
Here $z_0$ defines the position of the image plane typically located at 
$2$a$_0$ (a$_0$ stands for the Bohr radius) above the surface atomic layer \cite{Chulkov1999}. Without loss of generality, we set $z_0=0$ (see FIG.\,\ref{fig:WPPgeometry}).

\begin{itemize}
  \item $V_{\rm{mol}}(\vec{r})$ is the electron-molecule interaction potential. 
It is given by the sum of local $V^L(\vec{r})$ and nonlocal $V^{NL}(\vec{r})$ terms $V_{\rm{mol}}(\vec{r})=V^L(\vec{r})+V^{NL}(\vec{r})$. 
  \item $V_{\rm surf}(z)$ is the electron-metal surface interaction 
potential.
  \item $\Delta V_{\rm surf}(\vec{r})$ stands for the change of
$V_{\rm surf}(z)$ induced by the molecule.
\item $V_{\rm{abs}}(\vec{r})$ is the optical absorbing potential.
\end{itemize}

With time-dependent wave function $\psi{(\vec{r},t)}$ represented on the 
equidistant 3D Cartesian grid, Eq.~\ref{eq:MainWPP} is solved using the Fourier 
pseudo-spectral approach \cite{Kosloff1993} and the split-operator technique 
\cite{LEFORESTIER1991} (the time-propagation step is $dt = 0.02$~a.u.). The 3D Cartesian grid typically comprises $N_x=768, N_y=768, N_z=1280$ nodes and has the same step 
$h=0.11$~a$_0$ in $x$, $y$, and $z$-directions. 
Analysis of $\psi{(\vec{r},t)}$ yields the energies and electron transfer rates 
of the quasistationary molecular states in front of the metal surface.\\

Below we detail the components of the effective one-electron potential.

\textbf{The electron-molecule interaction potential}, $V_{\rm{mol}}(\vec{r})=V^L(\vec{r})+V^{NL}(\vec{r})$, is 
obtained from the ab-initio quantum chemistry Density Functional Theory (DFT) 
calculations as implemented in the Abinit package \cite{GONZE2002ABINIT}. 
The calculations are performed for the free standing molecule with 
the equilibrium geometry.

Within the one-electron approach to the RET between the molecule and the 
surface, it is important that the energies of the active orbitals are representative for the energies of the many-body states with respect to the vacuum level and the Fermi energy of the metal as it defines the height of the potential barrier and the direction of the electron transfer ("to" or "from" the metal). Since the molecular anion is formed by an electron attachment to the neutral molecule, the binding energy of the LUMO with respect to the vacuum level should represent the electron affinity of PTCDA ($\SI{3.25}{eV}$ \cite{Hofmann_2013}, $\SI{3.07}{eV}$ \cite{Haitao2016}). However, since Koopmans' theorem \cite{Koopmans1934} does not apply for the DFT-case, the binding energy of the LUMO as obtained with Abinit is appreciably 
different from the molecular affinity. We then use the conclusions of the study 
performed with hybrid functionals \cite{Hofmann_2013}, and defined 
the $V_{\rm{mol}}(\vec{r})$ potential as follows.
\begin{equation}\label{eq:hybridpot}
  V_{\rm{mol}}(\vec{r})=
  \underbrace{\zeta V^L_{\rm{PTCDA}}(\vec{r}) + (1-\zeta)~V^L_{\rm{PTCDA}^-}(\vec{r})}_{V^L(\vec{r})} +V^{NL}(\vec{r}),
\end{equation}
where $V^L_{\rm{PTCDA}}(\vec{r})$ is the local potential (the sum of Hartree and exchange-correlation contributions) calculated for the free-standing neutral molecule, and $V^L_{\rm{PTCDA}^-}(\vec{r})$ is the local potential calculated for the free-standing molecular anion assuming 1/2 occupation of the frontier orbitals of both spins. The nonlocal potential is given by the norm-conserving pseudopotentials in the Kleynman Bylander form \cite{Kleinman1982} as implemented in Abinit to describe the electron interaction with individual atoms forming the molecule. For $\zeta=0.32$ from the WPP calculations for the free-standing molecule, we obtain the energy of the LUMO $E_{\rm{LUMO}}=-\SI{3.19}{eV}$ with respect to the vacuum level, i.e., in good agreement with PTCDA affinity reported earlier \cite{Hofmann_2013,Haitao2016}. Simultaneously $E_{\rm{LUMO}+1}=-\SI{1.88}{eV}$ and $E_{\rm{LUMO}+2}=-\SI{1.85}{eV}$ so that $E_{\rm{LUMO}+2}-E_{\rm{LUMO}}=\SI{1.34}{eV}$ close to the trion fluorescence $X^-$ peak energies observed in our experiments.

\textbf{The electron-metal surface interaction} $V_{\rm surf}(z)$ is represented by the model potential of Jennings at al \cite{Jennings1988}. 
\begin{align}
\label{eq:SurfacePot}
    V_{\rm surf}(z)=\begin{cases} 
    -\frac{1}{4(z-z_0)} \left\{1-e^{-\alpha_{\rm J}(z-z_0)} \right\} , &z>z_0, \\ 
    -\frac{V_0}{A_{\rm J} e^{B_{\rm J}(z-z_0)}+1}, &\rm{otherwise}, \end{cases}
\end{align}
\\
where $A_{\rm J}=4V_0/\alpha_{\rm J}-1$, $B_{\rm J}=2V_0/\alpha_{\rm J}$. 
As follows from the equation above, $V_{\rm surf}(z)$ is only a function of the 
electron coordinate $z$ perpendicular to the surface. It smoothly joins 
the classical image potential $-1/4(z-z_0)$ for an electron being in a vacuum 
($z \gg z_0$) with a constant potential $-V_0$ inside metal. 
We have also performed the calculations with the
$z$-dependent model potential developed by Chulkov and collaborators 
\cite{Chulkov1999} to describe an electron interaction with Ag(111) metal 
surface corresponding to the substrate. In this case the band structure of 
Ag(111) along the direction perpendicular to the surface is well reproduced 
(projected band gap, surface and image potential states). Consistent with earlier 
results \cite{WPP_Method2021,FernandoAnisotropic2021}, the projected band 
structure shows only mild effect on the electron transfer process. 
Considering this result as well as the experimental procedure where 
the STM tip is covered with Ag layer by indentation of the tip into Ag 
surface so that the structure of the Ag layer is ill-defined, we 
decided to use the jellium description of the metal to characterise the 
electron transfer with both: the tip and the substrate. The potential 
parameters are set as $\alpha_{\rm J}=1.1715$\,a.u., $V_0=\SI{12}{eV}=0.4410$\,a.u. 
consistent with the potential tail at metal/vacuum interface and the valence 
band bottom of silver \cite{Chulkov1999}.

\textbf{The change of the electron-metal interaction 
because of the presence of the molecule} $\Delta V_{\rm{surf}}(\vec{r})$ is 
set as 
\begin{align}
\label{eq:MolecImage}
    \Delta V_{\rm{surf}}(\vec{r})=\begin{cases} 
    -V^L(x,y,-(z+a)), &z>0. \\ 
    0, &\rm{otherwise}\end{cases}
\end{align}
This choice guarantees that at the image potential plane defined with $\vec{r}_{\rm{IP}}=(x,y,0)$ 
the local part of the electron-molecule interaction is fully screened 
$V^L(\vec{r}_{\rm{IP}})+\Delta V_{\rm{surf}}(\vec{r}_{\rm{IP}})=0$.
Since the non-local potential $V^{NL}(\vec{r})$ consists of very 
short-range contributions around the atoms constituting the molecule (nonlocal 
potential range smaller than molecule-surface distance $a$), 
we also obtain that $V_{\rm{mol}}(\vec{r}_{\rm{IP}})
+\Delta V_{\rm{surf}}(\vec{r}_{\rm{IP}})=0$. The electron-molecule interaction is smoothly 
screened at the metal surface. 

\textbf{The optical absorbing potential} $V_{\rm{abs}}(\vec{r})$ 
is introduced at the boundaries of the computational box to impose the outgoing 
wave boundary conditions\cite{Riss_1993,Moiseyev_1998} consistent with the 
search for the quasi-stationary 
molecule-localized states coupled with the continuum of the electronic 
states of the metal. For the explicit form of $V_{\rm{abs}}(\vec{r})$ see 
Ref.~\onlinecite{WPP_Method2021}.

\subsection{Excitation mechanisms of suspended molecules}

\begin{figure}
  \includegraphics[width=0.9\linewidth]{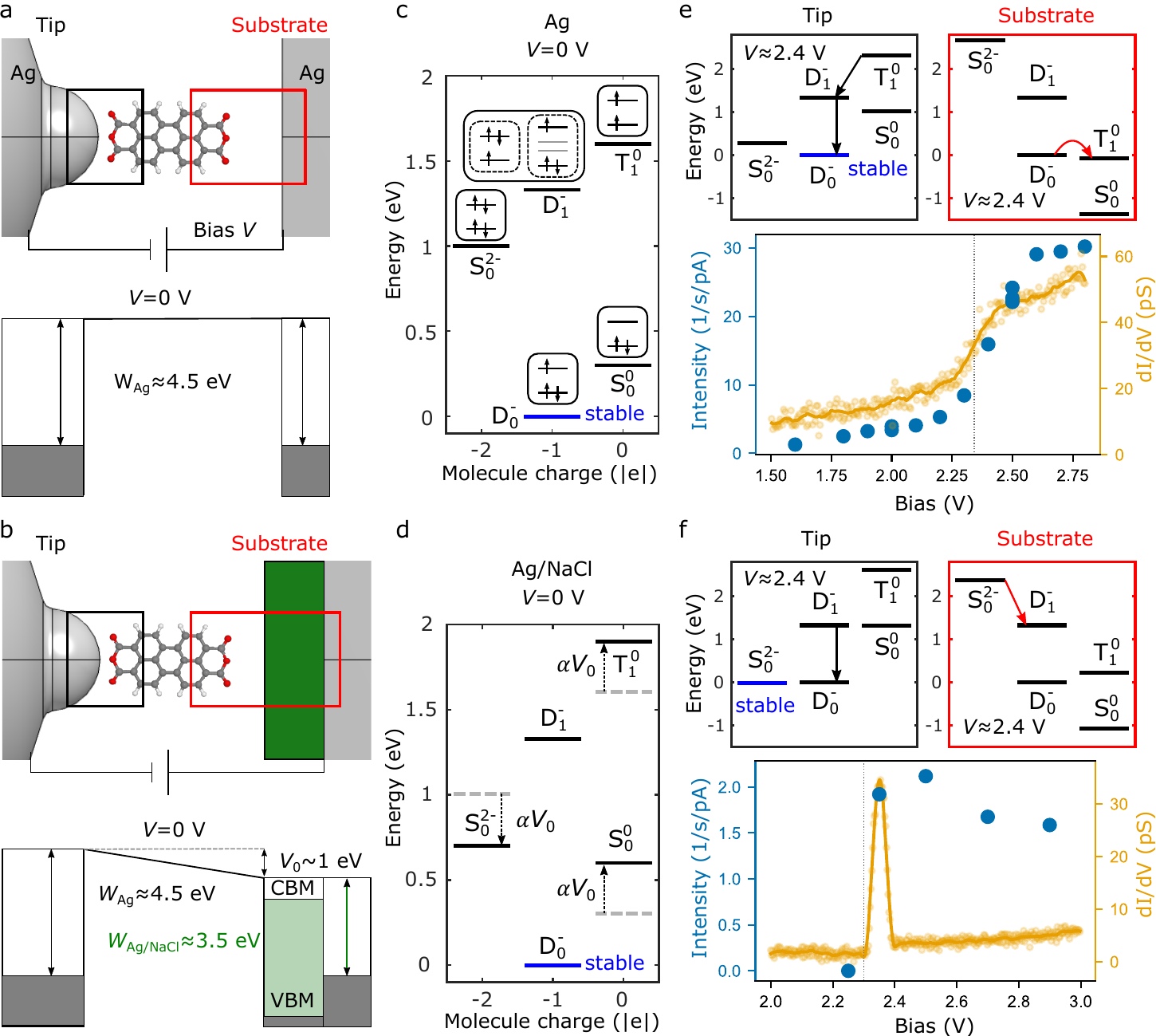}
  \caption{\label{fig:mechanisms} Schematic representation of the geometry of the molecule suspended on the tip and facing a \textbf{a} Ag(111) and \textbf{b} NaCl/Ag(111) surface. Alongside with the geometry we show schematic single-particle band diagrams characterising the tip and the substrate. In the diagram we mark the work functions $W_{\rm Ag}$ and $W_{\rm Ag/NaCl}$ as well as the offset voltage drop $V_0$ caused by the difference of the tip and the substrate work functions, and the conduction-band minimum (CBM) and valence-band maximum (VBM) of NaCl. In (c,d) we show many-body energy diagrams of the molecule where the charge is transferred from/to the Fermi level of the tip. The Ag substrate is considered in \textbf{c}, the NaCl/Ag in \textbf{d}. The state stabilized by the tip is drawn in blue and marked as stable. \textbf{e}, \textbf{f} $\didv$ curves, corresponding photon intensity, and many-body level diagrams drawn at the threshold voltage. In the black frame we show the diagrams assuming charge-exchange with the tip, and in the red frame the diagrams assuming charge exchange with the substrate.}
\end{figure}

In the main text, we briefly discuss the transport mechanism that drives the suspended PTCDA molecule in the excited state. Here, we provide some more details on these mechanisms, especially for cases where the suspended molecule is facing (i) a bare Ag(111) substrate and (ii) a Ag(111) substrate covered with 2ML of NaCl. In FIG.\,\ref{fig:mechanisms}\textbf{a} (top) we schematically show the geometry of the molecule suspended in the gap, and (bottom) a simplified band diagram of the Ag electrodes, the tip and the substrate, showing aligned work functions in the absence of external bias. In FIG.\,\ref{fig:mechanisms}\textbf{c} we show a many-body energy diagram of a PTCDA-tip facing a bare Ag(111). In this diagram we show the energies of the most relevant states that enter the charge transfer dynamics: The ground (excited) doublet state of the negative molecule D$_0^-$ (D$_1^-$), the neutral singlet ground state S$_0^0$, the neutral triplet state T$_1^0$, and the singlet ground state S$_0^{2-}$ of the doubly negative molecule. In the insets, we also schematically show the dominant electron configurations representing the respective states. Since the molecule is in electrical contact with the tip, we align the many-body levels with respect to the work function of the tip, as opposed to the representation where the levels are aligned with respect to the vacuum level. In this representation, it is therefore assumed that the molecule can exchange an electron with the tip and that this electron is at the Fermi level of the metal. Here, the lowest-lying state is naturally stabilized by the charge transfer with the tip.

The situation is somewhat different when the molecule is suspended in a gap between the tip and a NaCl/Ag(111) substrate as shown in FIG.\,\ref{fig:mechanisms}\textbf{b}(top). The NaCl layer is schematically depicted in the single-particle diagram in FIG.\,\ref{fig:mechanisms}\textbf{b}(bottom) via the green rectangle marking the gap between the valence band maximum - VBM - and the conduction band minimum - CBM. The work function of the NaCl/Ag(111) interface is significantly lowered compared to the Ag(111) surface. Since the Fermi levels of the tip and the substrate align at zero external bias, the difference of $\SI{1}{eV}$ between the vacuum level on the tip and the substrate side gives rise to a static electric field across the gap, which acts on the levels of the suspended molecule. This static field affects the level diagram of the suspended molecule as shown in FIG.\,\ref{fig:mechanisms}\textbf{d}. Here, the grey dashed lines mark the original positions of the levels shown in FIG.\,\ref{fig:mechanisms}\textbf{c}. These are shifted by the static electric field (expressed here as an effective bias offset of $V_0 = \SI{1}{V}$) towards higher energies for the neutral states, and to the lower energy for the doubly negative state. We assume that $\alpha=1/3$ of the voltage drop occurs on the tip side, which we approximately derive from the relative position of the molecule within the gap and which is consistent with the position of the tunneling thresholds in the $\didv$ spectra. 

We now examine what happens when an external bias is applied in the simpler case of a molecule facing the Ag(111) substrate. Here, the Fermi levels of the metal electrodes are no longer aligned and the many-body diagram of the PTCDA states depends on the electrode with which the molecule exchanges electrons. Moreover, the additional voltage drop between the tip and the molecule further modifies the alignment of the levels. 
In FIG.\,\ref{fig:mechanisms}\textbf{e} (top), we show such level diagrams drawn for PTCDA on top of the bare Ag(111). The black frame shows how the diagram changes upon the application of a voltage $V = \SI{2.4}{V}$, assuming that the molecule exchanges electrons with the tip. In contrast, the red frame shows the diagram assuming the same voltage conditions, but for electrons exchanged with the substrate. As the voltage drops on both sides of the molecule are different, this results in a different alignment of the many-body levels. A simple rule can be derived from the energetic ordering of the levels: a spontaneous exchange of an electron with the tip (the substrate) can occur only if the energy of the initial many-body state in the respective diagram is higher than the energy of the final one. This rule stems from the fact that an electron escaping the molecule must propagate into an unoccupied state of the metal above the Fermi level, and conversely, an electron captured by the molecule can only originate from the occupied states of the metal below the Fermi level. 
For the threshold bias of about $\SI{2.4}{V}$, we see that the molecule is stabilized by the efficient tip-mediated charge transfer in the state D$_0^-$, as can be seen in the diagram with the black frame. D$_0^-$ can with a smaller tunneling probability decay into S$_0^0$ and T$_1^0$ by releasing an electron into the substrate as shown in the red frame. On the occasion that T$_1^0$ is populated, the tip can rapidly provide an electron and bring the molecule to the D$_1^-$ excited state (black frame). This state can then radiatively decay back to the negative ground state D$_0^-$. These features are manifested in the $\didv$ spectra and corresponding photon-emission thresholds shown in FIG.\,\ref{fig:mechanisms}\textbf{e} (bottom). At the bias voltage of about $\SI{2.3}{V}$, we observe a smooth step in the $\didv$ curve, corresponding to the D$_0^-$ $\rightarrow$ T$_1^0$ transition, accompanied by a gradual onset of photon emission. 
The smooth onset of this process can be understood as a result of voltage-dependent tunneling rate between the molecule and the tip, a result of the modification of the tunneling barrier with increasing voltage. 

In contrast, for the molecule suspended above the 2ML-NaCl/Ag(111) substrate, we observe in the $\didv$ a sharp peak at $\sim \SI{2.35}{V}$ accompanied by a sharp onset of photon emission, as shown in FIG.\,\ref{fig:mechanisms}\textbf{f}(bottom). To understand this behavior, we plot in FIG.\,\ref{fig:mechanisms}\textbf{f}(top) the corresponding level diagrams at the bias of about $\SI{2.4}{V}$, roughly corresponding to the onset of the experimental features, aligned with respect to the tip (black box) and the substrate (red box). Because of the reduced barrier height on top of NaCl and the associated bias offset of the molecular energy levels, at $\sim \SI{2.35}{V}$ the molecule is stabilized in the doubly negative charge state S$_0^{2-}$ (black box). The sharp $\didv$ feature in FIG.\,\ref{fig:mechanisms}\textbf{f}(bottom) is consistent with such a charging event. From then, a substrate-mediated channel to $D_1^-$ opens (red box), eventually leading to the emission of the molecule suspended at the tip. In this example, the $D_1^-$ state is thus populated by a different mechanism than for the molecule-tip in front of the bare Ag(111).